\documentclass[preprint,12pt,number]{elsarticle}

\usepackage{amsmath}
\usepackage{amsfonts}
\usepackage{amssymb}
\usepackage{amsthm}
\usepackage[margin=1.5in]{geometry}
\usepackage{indentfirst}
\usepackage{booktabs}
\usepackage{chngcntr}
\usepackage{comment}
\usepackage{enumerate}
\usepackage{url}

\usepackage{algorithm}
\usepackage[noend]{algpseudocode}

\usepackage{graphicx}
\usepackage{tikz}
\usetikzlibrary{shapes,positioning}
\usepackage{caption}
\usepackage{subcaption}

\biboptions{numbers,sort&compress}
\usepackage{hyperref}

\counterwithout{remark}{section}
\counterwithout{figure}{section}
\counterwithout{table}{section}
\counterwithout{theorem}{section}
\counterwithout{lemma}{section}
\counterwithout{proposition}{section}

\begin{document}
\begin{frontmatter}

\title{Subgroup Inconsistency and the Dilution Effect in Levene’s Test for Homoscedasticity} 

\author{Vladimir Gurvich}
\ead{vgurvich@hse.ru and vladimir.gurvich@gmail.com}
\address{National Research University Higher School of Economics, Moscow, Russia\\ Rutgers Center for Operations Research, Rutgers University, Piscataway, New Jersey, United States}

\author{Mariya Naumova}
\ead{mnaumova@business.rutgers.edu}
\address{Rutgers Business School, Rutgers University, Piscataway, New Jersey, United States}

\begin{abstract}

\noindent

Levene's test for homoscedasticity is a standard procedure used to evaluate whether multiple groups of independent observations share a common variance. Because Levene's test relies on an Analysis of Variance (ANOVA) applied to transformed absolute or squared deviations, it structurally mirrors the statistical properties of ANOVA itself. Let $\mu_j$ and $\sigma^2_j$ denote the expectation and variance of the observations in group $j$. It was previously established that for a given significance level $\alpha$, ANOVA can result in a logical contradiction: failing to reject the global null hypothesis $H_0: \mu_1 = \mu_2 = \mu_3$ while simultaneously rejecting the localized hypothesis $H_0': \mu_1 = \mu_2$ with the same or higher confidence. In this paper, we show that Levene’s test directly inherits this same ``paradox'' regarding group variances $\sigma^2_j$. We provide theoretical reasoning and a numerical illustration of this inconsistency, demonstrating how the addition of a well-behaved third group can dilute the test statistic and mask a significant localized variance discrepancy.

\vspace{1em}

\end{abstract}


\end{frontmatter}

\section{Introduction}

Homoscedasticity — the assumption that multiple groups or populations share a common variance — is a foundational prerequisite for many classical statistical procedures, including ordinary least squares (OLS) regression and Analysis of Variance (ANOVA). When the assumption of equal variances is violated, standard inferential tests can suffer from severely inflated Type I error rates or compromised statistical power \cite{BB04, GGM09}. Consequently, formal diagnostic testing for variance homogeneity remains a standard preliminary step in applied quantitative research.

Early classical procedures for evaluating variance equality, such as Bartlett's test or the standard two-sample $F$-test, rely heavily on the assumption of multivariate normality. As demonstrated by Box \cite{B53} and Box and Watson \cite{BW62}, these classical variance tests are notoriously sensitive to distributional departures, frequently confounding non-normality (such as excess kurtosis) with true variance heterogeneity. To mitigate this fragility, Levene \cite{L60} introduced a robust alternative that applies a standard one-way ANOVA $F$-test to transformed absolute deviations of observations from their group means. Brown and Forsythe \cite{BF74} subsequently refined this procedure by substituting group medians for group means, significantly improving test performance in the presence of skewed or heavy-tailed distributions. Today, Levene's test and its median-based variant represent the prevailing benchmark for assessing homoscedasticity across biostatistics, econometrics, and applied operations research \cite{GGM09, P00}.

Despite its widespread adoption, Levene's procedure is fundamentally a statistical reduction: it transforms a hypothesis regarding population variances ($\sigma_1^2 = \dots = \sigma_k^2$) into an equivalent ANOVA model evaluating ``average absolute deviations.'' Consequently, any structural vulnerabilities or logical paradoxes inherent to the ANOVA framework are directly inherited by Levene's test. Gurvich and Naumova \cite{GN21} established that one-way ANOVA displays a logical non-monotonicity (or ``subgroup inconsistency'') when evaluating $k \ge 3$ groups. Specifically, an omnibus test can fail to reject the global null hypothesis $H_0: \mu_1 = \dots = \mu_k$ while a pairwise comparison simultaneously rejects $H_0': \mu_i = \mu_j$ for a specific pair $(i, j)$ at the same or higher confidence level. This counterintuitive behavior occurs because adding ``well-behaved'' intermediate groups disproportionately increases the error degrees of freedom and stabilizes the pooled within-group variance, thereby masking severe localized discrepancies.

In this paper, we demonstrate that Levene's test suffers from this exact ``dilution paradox.'' We show that for $k \ge 3$ populations, the addition of a third group with well-aligned sample deviations can artificially inflate the denominator of the Levene statistic $W$, causing the omnibus test to accept global variance homogeneity even when a localized variance discrepancy between two groups is statistically significant. It is important to emphasize that failing to reject the null hypothesis in Levene's test does not actually prove that the variances in all groups are equal. It merely indicates that there is insufficient statistical evidence to reject the null hypothesis of equality at a specified confidence level (e.g., $1 - \alpha$).

The remainder of this paper is structured as follows. Sections 2 and 3 formalize the Levene test statistic and detail its reduction to the standard one-way ANOVA linear model. Section 4 presents a numerical illustration demonstrating how a localized variance difference ($p = 0.0305$) is masked ($p = 0.0987$) upon introducing a third sample drawn from an identically distributed population. Section 5 establishes the theoretical framework for generating large infinite families of multi-group samples exhibiting subgroup inconsistencies for any $k \ge 3$. Section 6 compares the structural properties and sensitivity of Levene's test against the classical two-sample $F$-test. Finally, Section 7 provides concluding remarks on the methodological implications of statistical reductions.

\section{Levene test for homoscedasticity}

The Levene test provides a formal procedure for assessing the null hypothesis that $k$ samples originate from populations with equal variances, denoted as $H_0: \sigma_1^2 = \sigma_2^2 = \dots = \sigma_k^2$. Given a set of observations $Y_{ij}$, where $i = 1, \dots, k$ represents the group index and $j = 1, \dots, N_i$ represents the observation index within the $i$-th group, the test statistic $W$ is constructed by performing a one-way analysis of variance on the absolute deviations of the observations from their respective group central tendencies. Let $N = \sum N_i$ denote the total sample size. We define the transformed variables $Z_{ij} = |Y_{ij} - \bar{Y}_{i\cdot}|$, where $\bar{Y}_{i\cdot}$ is the arithmetic mean of the $i$-th group. The test statistic is then formulated as:

\begin{equation}
W = \frac{(N-k)}{(k-1)} \frac{\sum_{i=1}^k N_i (\bar{Z}_{i\cdot} - \bar{Z}_{\cdot\cdot})^2}{\sum_{i=1}^k \sum_{j=1}^{N_i} (Z_{ij} - \bar{Z}_{i\cdot})^2}
\end{equation}

In this expression, $\bar{Z}_{i\cdot} = \frac{1}{N_i} \sum_{j=1}^{N_i} Z_{ij}$ constitutes the group mean of the absolute deviations, and $\bar{Z}_{\cdot\cdot} = \frac{1}{N} \sum_{i=1}^k \sum_{j=1}^{N_i} Z_{ij}$ represents the grand mean of all deviations. Under the null hypothesis of homoscedasticity, and assuming the underlying distributions are sufficiently well-behaved, the statistic $W$ approximately follows an $F$-distribution with $d_1 = k - 1$ and $d_2 = N - k$ degrees of freedom. While the original formulation by Levene \cite{L60} utilized the group mean for the centering of $Y_{ij}$, subsequent refinements by Brown and Forsythe \cite{BF74} demonstrated that replacing the mean with the group median, $\tilde{Y}_{i\cdot}$, enhances the robustness of the test against heavy-tailed distributions and departures from normality. The structural integrity of $W$ as a ratio of the between-group variance to the within-group variance of the absolute deviations ensures that the test remains an effective diagnostic for variance heterogeneity in diverse experimental designs.

\section{Relation of the Levene Test to the ANOVA framework}

The Levene test statistic $W$ is mathematically equivalent to the $F$-statistic generated from a one-way Analysis of Variance (ANOVA) performed on the transformed absolute deviations $Z_{ij}$. To formalize this relationship, we define $Z_{ij}$ by centering the observations $Y_{ij}$ around a measure of group central tendency, denoted as $C_i$:

\begin{equation}
Z_{ij} = |Y_{ij} - C_i|
\end{equation}

In the original formulation proposed by Levene, $C_i = \bar{Y}_{i\cdot}$ (the group arithmetic mean) \cite{L60}. In the more robust Brown--Forsythe variation, $C_i = \tilde{Y}_{i\cdot}$ (the group median) \cite{BF74}. The resulting values $Z_{ij}$ are modeled using the linear structure of a fixed-effects one-way ANOVA:

\begin{equation}
Z_{ij} = \mu_Z + \alpha_i + \epsilon_{ij}
\end{equation}
where $\mu_Z$ represents the grand mean of the absolute deviations, $\alpha_i$ denotes the effect of the $i$-th group on the magnitude of deviation, and $\epsilon_{ij}$ is the random error term. Under this framework, the null hypothesis of equal variances $H_0: \sigma_1^2 = \sigma_2^2 = \dots = \sigma_k^2$ is mapped onto the null hypothesis of equal ``deviation means'' across groups:

\begin{equation}
H_0: \alpha_1 = \alpha_2 = \dots = \alpha_k = 0
\end{equation}

The statistic $W$ is derived as the ratio of the Mean Square Between groups ($MSB_Z$) to the Mean Square Within groups ($MSW_Z$) of these transformed values:

\begin{equation}
W = \frac{MSB_Z}{MSW_Z} = \frac{\frac{1}{k-1} \sum_{i=1}^k N_i (\bar{Z}_{i\cdot} - \bar{Z}_{\cdot\cdot})^2}{\frac{1}{N-k} \sum_{i=1}^k \sum_{j=1}^{N_i} (Z_{ij} - \bar{Z}_{i\cdot})^2}
\end{equation}

Thus, the Levene test utilizes the standard ANOVA machinery to detect differences in the ``average distance from the center'' across groups. A statistically significant $W$ indicates that the average dispersion (and thus the variance) differs significantly among the $k$ populations.

\section{Numerical example}
We present a small example with three generated samples from normal distributions with means $\mu_1 = \mu_2 = \mu_3 = 0$ and variances $\sigma_1^2 = \sigma_2^2 = \sigma_3^2= 4$; see Table 1.

\begin{table}[h!]
\centering
\caption{Generated independent samples from normal distributions with means $\mu_1 = \mu_2 = \mu_3 = 0$ and variances $\sigma_1^2 = \sigma_2^2 = \sigma_3^2= 4$, $N_1 = N_2 = N_3 = 15$.}
\small
\begin{tabular}{crrr}
\toprule
Observation & Group 1 & Group 2 & Group 3\\
\midrule
1  & -0.8335 & -0.9126 & -1.4185 \\
2  &  0.1875 &  0.2356 &  0.4402 \\
3  & -3.9724 & -0.7470 &  0.0125 \\
4  &  3.5806 & -1.5643 &  0.2045 \\
5  & -3.2929 & -1.9182 &  0.3330 \\
6  &  1.1105 &  1.7547 & -0.4064 \\
7  &  2.4001 &  0.3385 &  2.1051 \\
8  & -1.1436 &  0.0722 &  0.4862 \\
9  & -1.9464 & -1.9924 & -1.6589 \\
10 &  0.4213 &  0.0298 &  1.1656 \\
11 & -1.9144 & -1.4529 & -1.0433 \\
12 &  1.5482 & -0.8918 & -2.2510 \\
13 & -0.5627 & -0.3685 &  1.3986 \\
14 &  1.3833 &  1.2982 &  2.3940 \\
15 & -0.5170 & -0.1325 & -1.3781 \\
\bottomrule
\end{tabular}
\end{table}

The spread of the three groups is demonstrated in Figure 1. Note how Group 3 provides a ``bridge'' between the highly variable Group 1 and the more compact Group 2.

\begin{figure}[h!]
    \centering
    \begin{subfigure}[b]{0.48\textwidth}
        \centering
        \includegraphics[width=\textwidth]{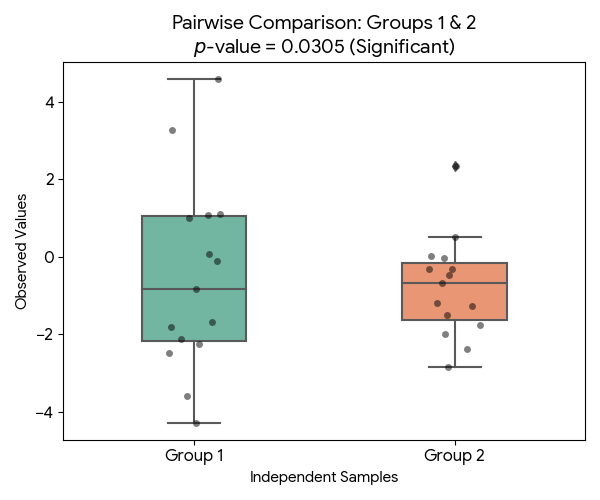}
        \caption{Pairwise comparison of Group 1 and Group 2. The high variation in Group 1 leads to rejection of $H_0$ ($p=0.0305$).}
        \label{fig:two_samples}
    \end{subfigure}
    \hfill 
    \begin{subfigure}[b]{0.48\textwidth}
        \centering
        \includegraphics[width=\textwidth]{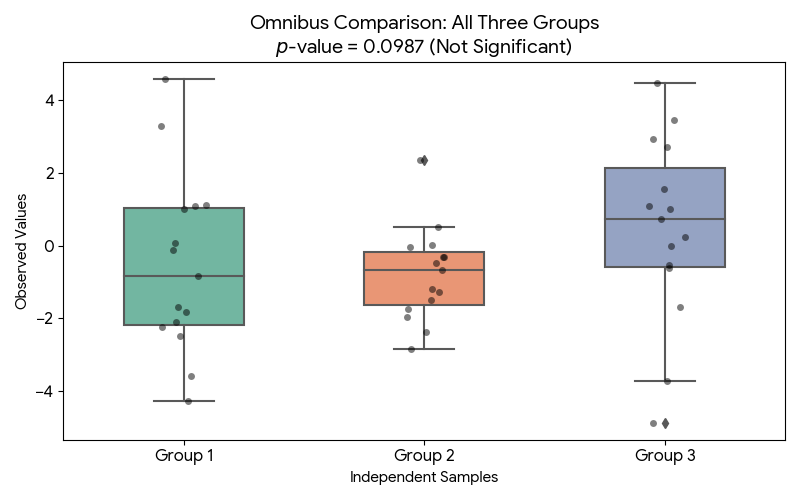}
        \caption{Omnibus comparison of all three groups. The introduction of Group 3 stabilizes the denominator, resulting in a failure to reject $H_0$ ($p=0.0987$).}
        \label{fig:three_samples}
    \end{subfigure}
    
    \caption{Visual comparison of sample distributions demonstrating the dilution effect in Levene's test. (a) A pairwise comparison of Group 1 and Group 2, highlighting the significant localized discrepancy in their variances. (b) An omnibus view including Group 3. The addition of Group 3 provides a centralized distribution of deviations, which stabilizes the pooled within-group variance and increases the degrees of freedom, ultimately masking the variance discrepancy between Groups 1 and 2.}
    \label{fig:main_comparison}
\end{figure}

We evaluate the equality of variances across different subsets of the data at the $\alpha = 0.05$ significance level. Table 2 summarizes the shift in the test statistic and $p$-value upon the inclusion of the third sample.

\begin{table}[h]
\centering
\caption{Comparison of Levene's Test results ($k=2$ vs. $k=3$)}
\begin{tabular}{lcccc}
\toprule
Null hypothesis ($H_0$) & Statistic ($W$) & $p$-value & Decision \\
\midrule
$\sigma_1^2 = \sigma_2^2$ & 5.161 & 0.0305 & Reject $H_0$ \\
$\sigma_1^2 = \sigma_2^2 = \sigma_3^2$ & 2.449 & 0.0987 & Fail to reject $H_0$ \\
\bottomrule
\end{tabular}
\end{table}

The transition from a significant result in the pairwise case to a non-significant result in the omnibus case illustrates the non-monotonic nature of the $p$-value in this context. Despite the theoretical variance being constant across all populations, the realized sample variance of Group 3 serves to mitigate the localized discrepancy between Groups 1 and 2.

This outcome illustrates the sensitivity of the $F$-distribution's denominator to ``well-behaved'' data. The introduction of Group 3, despite having the same theoretical variance, contributed a distribution of absolute deviations that aligned closely with the global median of deviations. This increased the error degrees of freedom and smoothed the variance estimate, ultimately ``diluting'' the localized discrepancy between Groups 1 and 2. Researchers should be cautious of concluding variance homogeneity solely based on the addition of supplemental groups.

\section{Large infinite families of samples with $k$ groups}

The numerical example provided in Table 1 is not an isolated incident; rather, it represents a specific instance of a broader structural property of the Levene test. Since the Levene statistic $W$ is mathematically equivalent to an ANOVA $F$-test performed on the absolute deviations $Z_{ij} = |Y_{ij} - C_i|$, we can apply the geometric and probabilistic arguments from \cite{GN21} to construct large infinite pseudo-random families of samples that exhibit logical contradictions for any $k \geq 3$.

\subsection{Construction of Case (i) for the Levene Test}

In Case (i), the omnibus Levene test rejects the global null hypothesis $H_0^L: \sigma_1^2 = \dots = \sigma_k^2$, while no pairwise Levene test rejects $H_0^L(i,j): \sigma_i^2 = \sigma_j^2$ for any $i, j \in \{1, \dots, k\}$. 

To construct such a family, let the absolute deviations $Z_{ij}$ be generated such that the group means $\bar{Z}_{i\cdot}$ are distributed widely enough to make the between-group variance $MSB_Z$ large, yet the distance between any specific pair $|\bar{Z}_{i\cdot} - \bar{Z}_{j\cdot}|$ remains below the critical threshold for the pairwise $F$-test (or the studentized range critical value). 

For a fixed significance level $\alpha$ and total sample size $N$, we can define a region in the $(k-1)$-dimensional space of group means $\bar{Z}_{i\cdot}$ where the omnibus $W$ statistic exceeds $F_{\alpha, k-1, N-k}$, but all pairwise differences are statistically insignificant. As $k$ increases, the volume of this "contradiction region" in the parameter space of sample variances grows, allowing for the generation of infinite sequences of samples $\{Y_{ij}\}$ satisfying these conditions.

\subsection{Construction of Case (ii) for the Levene Test}

Case (ii) presents the more severe logical contradiction: a specific pair $(i,j)$ is found to have significantly different variances, yet the omnibus test fails to detect any heteroscedasticity across the $k$ groups. 

Consider a family of samples where Groups 1 and 2 have a large discrepancy in their average absolute deviations, $| \bar{Z}_{1\cdot} - \bar{Z}_{2\cdot} | > \Delta_{crit}$. We then introduce $k-2$ "buffer" or "bridge" groups. If these additional groups have:
\begin{enumerate}[(a)]
    \item Sample mean deviations $\bar{Z}_{m\cdot}$ (for $m=3, \dots, k$) that are close to the grand mean $\bar{Z}_{\cdot\cdot}$, and
    \item Sufficiently large within-group sample sizes $N_m$ or small internal variances $S^2_{Z,m}$,
\end{enumerate}
then the denominator of the omnibus statistic (the pooled $MSW_Z$) increases its degrees of freedom while the numerator's growth is stunted by the central positioning of the additional groups. 

Mathematically, for any $\alpha$, there exists a $k$ large enough such that the ``dilution" provided by the $k-2$ groups reduces the omnibus $W$ below the critical $F$-value, even though the restricted statistic $W_{1,2}$ remains significant. This demonstrates that the Levene test is not monotonic with respect to the inclusion of additional populations, leading to the conclusion that $\sigma_1^2 = \dots = \sigma_k^2$ is accepted even when $\sigma_1^2 = \sigma_2^2$ is rejected.

\subsection{Pseudo-random Generation and the Gauss-Markov Analogy}

For these constructions, we assume the transformed variables $Z_{ij}$ are approximately normal, which is justified for large $N_i$ by the central limit theorem, or by selecting specific parent distributions for $Y_{ij}$ (such as folded normal distributions). While it is difficult to verify the Gauss-Markov assumptions for the Multivariable Linear Regression (MLR) extension of the Levene test, the identity of the $F$-distribution ensures that these families of contradictions exist as a matter of algebraic necessity within the ANOVA framework.

\section{Comparison of Levene's Test and the Standard $F$-Test}

The assessment of homoscedasticity for the case of $k=2$ samples may be conducted using either the Levene test or the traditional $F$-test for the equality of two variances. However, these procedures are fundamentally distinct in their mathematical construction and their sensitivity to underlying distributional assumptions. The standard $F$-test is defined by the ratio of the sample variances:

\begin{equation}
F = \frac{s_1^2}{s_2^2} = \frac{\frac{1}{N_1 - 1} \sum_{j=1}^{N_1} (Y_{1j} - \bar{Y}_{1\cdot})^2}{\frac{1}{N_2 - 1} \sum_{j=1}^{N_2} (Y_{2j} - \bar{Y}_{2\cdot})^2}
\end{equation}

This test statistic $F$ follows an $F$-distribution with $N_1 - 1$ and $N_2 - 1$ degrees of freedom under the strict null hypothesis that the populations are normally distributed. It is well-documented in statistical literature that the $F$-test is highly non-robust to departures from normality \cite{B53, BW62, BB04}; specifically, it is sensitive to the kurtosis of the parent distributions. In contrast, the Levene test statistic $W$ operates on the absolute deviations $Z_{ij} = |Y_{ij} - \bar{Y}_{i\cdot}|$. By effectively performing a one-way ANOVA on these deviations rather than a ratio of squared residuals, the Levene test provides a more stable inference when the normality assumption is violated \cite{GGM09}.

Empirical studies indicate that while the $F$-test may provide slightly higher power under perfect normality \cite{M74, P00}, the Levene test, particularly the median-centered Brown-Forsythe variant \cite{BF74}, is superior in controlling the Type I error rate across a broad range of non-normal distributions. Consequently, for $k=2$, the two tests will generally yield different $p$-values, and the Levene test is typically preferred for its structural robustness in practical applications \cite{L60}. Further extensions of these tests continue to address specific sampling conditions, such as the inclusion of paired observations \cite{DRTW18}.

\section{Conclusion}
This research contributes to a broader program of critical analysis regarding the logical foundations of classical statistical testing. While recent work has exposed similar inconsistencies in Pearson’s chi-square test for homogeneity \cite{GN25} — specifically its lack of invariance under scaling — the present study focuses on the inheritance of structural flaws through statistical reduction. By demonstrating that Levene's test is susceptible to the same subset paradox previously identified in the one-way ANOVA, we show that these logical contradictions are not isolated to mean-based comparisons but extend to the fundamental assessment of homoscedasticity. Together, these findings suggest a systemic vulnerability in traditional frequentist methodologies where mathematically valid procedures can yield logically incompatible results.

\end{document}